# Wicked Implications for Human Interaction with IoT Sensor Data

Albrecht Kurze

Chemnitz University of Technology, Albrecht.Kurze@informatik.tu-chemnitz.de

Andreas Bischof

Chemnitz University of Technology, Andreas.Bischof@informatik.tu-chemnitz.de

Human data interaction with sensor data from smart homes can cause some implications when it comes to human sensemaking of this data. With our data-driven method *Guess the Data* for individual and collective data work we revealed in previous work a number of potential pitfalls when interacting with this type of data. We introduce some of the identified, often wicked implications for further discussion.

## 1 INTRODUCTION

Smart Home Internet of Things (IoT) technologies increasingly collect seemingly inconspicuous data from homes using simple sensors, e.g. for temperature or light etc. Prior work has shown that not only automated data analytics [7,8] but also human sensemaking of such sensor data, as part of human data interaction, can reveal domestic activities [3,10]. Our research using our method *Guess the Data* [6] furthermore showed that this ability is not limited to experts or members of a household but also possible through using 'situated knowledge' [4] for others, e.g. friends and neighbors etc. [1,6,9]. We were able to show this through extending the concept of 'data work' [2,3]. Our participants not only performed individual data work at home on their own data, but also collective data work on the anonymized data of others in group discussions. We also revealed risks of uses of senor data, i.e. privacy intrusions and lateral surveillance. In terms of collecting, storing, and processing sensor data from the home as well as interacting with it, such implications always have to be considered, from ideating a data-driven IoT solution throughout the complete design process [5]. We would like to discuss these implications and how we as HCI community can address them in human data interaction.

## 2 WICKED IMPLICATIONS

*Guess the Data* revealed a number of wicked implication and potential for fails when it comes to human data interaction. Sensor data in the home and the interaction with this data are not free of implications, **even if …**

…**no complex sensors** like cameras/microphones are involved: Even simple sensors can reveal a lot about domestic activities and behavior of inhabitants as our own and other findings show [3,10].

**… no AI/ML** is involved: People can interpret some simple sensor data on their own, at least in a fundamental way, even if it is only in a simple visualization like a line graph etc. *Guess the Data* underlines that simple sensor data and other data (personal and context) are linked to situated knowledge for interpretation. Our findings confirm that the interpretation about sensor data from home heavily relies on the specific context and the social order, as other studies have already shown [3,10].

**…no Big Brother** is involved: Within a close social context such as a household the situatedness is given for the members. This puts into perspective the assumption that bare data from simple sensors is harmless for privacy, since the gap for interpretation is supposedly too big. As we have seen that even supposed activities can lead to moral justification issues. The less abstract and more concrete threat for an individual's privacy might not come from an anonymous third party Big Brother but a big mother or other member of the household.

**… no family** or other household members are involved: Our findings show that even a little situated knowledge turns a guess on the data into an educated guess. In conclusion it is not necessary to be a member of a household to generate a coarse assumption with at least some plausibility (extending/contrasting [3]). We have seen this in our own data work in a preparation of a discussion as well as in the collective data work of the participants. In terms of implications, this might become critical because the interpretations lead to topics, which could be easily exploited by third parties.

**… no objective true interpretation** is given or possible: Even an educated guess can be objectively wrong. An assumption-based interpretation of the sensor data just needs to be plausible in means of situatedness and in most cases it is neither verifiable nor falsifiable based solely on the simple sensor data. In using sensor data as evidence, data becomes subjective truth. In believing the interpretation it becomes intersubjective truth. This way the validity of a subjective human interpretation is inflated and universalized.

… **no personal data** is directly involved: Even simple sensor data becomes personal data in just the moment when it is possible to relate this data to a specific person. Whether this need to objectively true or just (inter)subjectively plausible is another question that needs consideration. This implication directly leads to legal objections in the same way as for complex sensors such as cameras or microphones and interaction with their data.

**… no cloud** is used: The implications even emerge in a local data domain of a household or family as soon as data is accessible or shared. Even within the close social context of a family as a data domain the access to raw data has to be questioned. Appropriate mechanisms for collecting, storing, processing, and accessing of the sensor data as well as interacting with it are needed.

**… no evil** is intended: An implication of the interpretability of sensor data within situated knowledge is that once the data is there and shared, we have to expect uses and misuses. The participative approach in our method allowed observing how the sensor data in the home led to ethically problematic behavior within days. Regarding social consequences, we found examples where data was used for moral accountability, moral judgements, to encourage or even enforce behavior changes and for surveillance. An asymmetry of power between those having the ability to interact with this data and those having no access fosters the misuse. Having power over the sensors encourages the more powerful to (mis)use the data for their purposes, from subtle influencing to extortion.

## ACKNOWLEDGMENTS

This research is funded by the German Ministry of Education and Research (BMBF), grant FKZ 16SV7116.